\documentclass[reprint, superscriptaddress,showpacs,amsmath, amssymb, aps,prl,]{revtex4-1}

\usepackage{color}
\usepackage{ulem}

\usepackage{natbib}
\usepackage{amsmath}
\usepackage{siunitx}
\usepackage{graphicx}
\usepackage{subfigure}
\usepackage{multirow}
\usepackage{dcolumn}
\usepackage{bm}

\usepackage{hyperref}

\newcommand{\RNum}[1]{\uppercase\expandafter{\romannumeral #1\relax}}
\hypersetup{colorlinks=true, citecolor=blue, urlcolor=blue, linkcolor=blue}

\fontsize{3.0pt}{\baselineskip}\selectfont

\begin{document}
	\title{Nonlinear Enhancement of Measurement Precision  via a Hybrid Quantum Switch
    }

 \author{Lei Chen}

\affiliation{CAS Key Laboratory of Quantum Information, University of Science and Technology of China, Hefei 230026, People's Republic of China}
\affiliation{CAS Center For Excellence in Quantum Information and Quantum Physics, University of Science and Technology of China, Hefei, Anhui 230026, China}
\affiliation{Anhui Province Key Laboratory of Quantum Network, Hefei, Anhui 230026, China.}

\author{Yuxiang Yang}
\affiliation{QICI Quantum Information and Computation Initiative, Department of Computer Science, The University of Hong Kong,
Pokfulam Road, Hong Kong 999077, China}

\author{Gong-Chu Li}
\affiliation{CAS Key Laboratory of Quantum Information, University of Science and Technology of China, Hefei 230026, People's Republic of China}
\affiliation{CAS Center For Excellence in Quantum Information and Quantum Physics, University of Science and Technology of China, Hefei, Anhui 230026, China}
\affiliation{Anhui Province Key Laboratory of Quantum Network, Hefei, Anhui 230026, China.}
\affiliation{Hefei National Laboratory, University of Science and Technology of China, Hefei, China}

\author{Xu-Song Hong}
\affiliation{CAS Key Laboratory of Quantum Information, University of Science and Technology of China, Hefei 230026, People's Republic of China}
\affiliation{CAS Center For Excellence in Quantum Information and Quantum Physics, University of Science and Technology of China, Hefei, Anhui 230026, China}
\affiliation{Anhui Province Key Laboratory of Quantum Network, Hefei, Anhui 230026, China.}
\affiliation{Hefei National Laboratory, University of Science and Technology of China, Hefei, China}

\author{Si-Qi Zhang}
\affiliation{CAS Key Laboratory of Quantum Information, University of Science and Technology of China, Hefei 230026, People's Republic of China}
\affiliation{CAS Center For Excellence in Quantum Information and Quantum Physics, University of Science and Technology of China, Hefei, Anhui 230026, China}
\affiliation{Anhui Province Key Laboratory of Quantum Network, Hefei, Anhui 230026, China.}

\author{Hua-Qin Xu}
\affiliation{CAS Key Laboratory of Quantum Information, University of Science and Technology of China, Hefei 230026, People's Republic of China}
\affiliation{CAS Center For Excellence in Quantum Information and Quantum Physics, University of Science and Technology of China, Hefei, Anhui 230026, China}
\affiliation{Anhui Province Key Laboratory of Quantum Network, Hefei, Anhui 230026, China.}

\author{Yuan-Cheng Liu}
\affiliation{CAS Key Laboratory of Quantum Information, University of Science and Technology of China, Hefei 230026, People's Republic of China}
\affiliation{CAS Center For Excellence in Quantum Information and Quantum Physics, University of Science and Technology of China, Hefei, Anhui 230026, China}
\affiliation{Anhui Province Key Laboratory of Quantum Network, Hefei, Anhui 230026, China.}

\author{Giulio Chiribella}
\email{giulio@cs.hku.hk}
\affiliation{QICI Quantum Information and Computation Initiative, Department of Computer Science, The University of Hong Kong,
Pokfulam Road, Hong Kong 999077, China}
\affiliation{Department of Computer Science, University of Oxford, Parks Road, Oxford OX1 3QD, United Kingdom}
\affiliation{Perimeter Institute for Theoretical Physics, Caroline Street, Waterloo, Ontario N2L 2Y5, Canada}

\author{Geng Chen}
\email{chengeng@ustc.edu.cn}
\affiliation{CAS Key Laboratory of Quantum Information, University of Science and Technology of China, Hefei 230026, People's Republic of China}
\affiliation{CAS Center For Excellence in Quantum Information and Quantum Physics, University of Science and Technology of China, Hefei, Anhui 230026, China}
\affiliation{Anhui Province Key Laboratory of Quantum Network, Hefei, Anhui 230026, China.}
\affiliation{Hefei National Laboratory, University of Science and Technology of China, Hefei, China}

\author{Chuan-Feng Li}
\email{cfli@ustc.edu.cn}
\affiliation{CAS Key Laboratory of Quantum Information, University of Science and Technology of China, Hefei 230026, People's Republic of China}
\affiliation{CAS Center For Excellence in Quantum Information and Quantum Physics, University of Science and Technology of China, Hefei, Anhui 230026, China}
\affiliation{Anhui Province Key Laboratory of Quantum Network, Hefei, Anhui 230026, China.}
\affiliation{Hefei National Laboratory, University of Science and Technology of China, Hefei, China}

\author{Guang-Can Guo}
\affiliation{CAS Key Laboratory of Quantum Information, University of Science and Technology of China, Hefei 230026, People's Republic of China}
\affiliation{CAS Center For Excellence in Quantum Information and Quantum Physics, University of Science and Technology of China, Hefei, Anhui 230026, China}
\affiliation{Anhui Province Key Laboratory of Quantum Network, Hefei, Anhui 230026, China.}
\affiliation{Hefei National Laboratory, University of Science and Technology of China, Hefei, China}

\begin{abstract}
Quantum metrology promises measurement precision beyond the classical limit by using suitably tailored quantum states and detection strategies. However, scaling up this advantage is experimentally challenging, due to the difficulty of generating high-quality large-scale probes.  
Here, we build a photonic setup that achieves enhanced precision scaling by manipulating the probe's dynamics through operations performed in a coherently controlled order.   Our setup applies an unknown rotation and a known orbital angular momentum increase in a coherently controlled order, in a way that reproduces a hybrid quantum SWITCH involving gates generated by both discrete and continuous variables.  
The unknown rotation angle $\theta$ is measured with precision scaling as $1/4ml$ when a photon undergoes a rotation of $2m\theta$ and an angular momentum shift of  $2l \hbar$. 
With a practical enhancement factor as high as 2317, the ultimate precision in our experiment is $0.0105^{\prime \prime}$ when using $7.16\times10^7$ photons, corresponding to a normalized precision of $\approx 10^{-4}$rad per photon. No photon interaction occurs in our experiment, and the precision enhancement consumes only a linearly increasing amount of physical resources while achieving a nonlinear scaling of the precision. We further indicate that this nonlinear enhancement roots in an in-depth exploration of the Heisenberg uncertainty principle (HUP), and our findings not only deepen the understanding of the HUP but also pave a pathway for advancements in quantum metrology.
\end{abstract}
\maketitle

\section{INTRODUCTION}
Precision measurements normally consist of three steps: preparation of the probe state,  encoding of the target parameter, and detection. Standard methods using  $N$ unentangled probe particles can reduce the measurement uncertainty to  $\Delta\propto 1/\sqrt{N}$, corresponding to the central limit scaling of classical statistics \cite{giovannetti2004quantum,giovannetti2006quantum}.  
Quantum metrology offers an opportunity to boost the precision beyond the classical limit, for example by initializing the probe particles into suitably designed quantum states, such as  NOON states \cite{bollinger1996optimal,afek2010high,nagata2007beating} or squeezed states \cite{grangier1987squeezed}. With such quantum resources as the probe, the precision can surpass the classical limit or even attain the Heisenberg limit  $\Delta \propto 1/N$ \cite{zwierz2010general}. To date, however, preparing large-scale quantum entangled states \cite {thomas2022efficient,wang201818} is still a daunting task, and hence the $1/N$ scaling can only survive for relatively small values of $N$. The squeezing-enhanced precision is feasible within the scope of current technological capabilities. However, the enhancement factor is constrained and highly susceptible to noise, such as photon loss \cite{demkowicz2013fundamental}. Although it has been shown that the Heisenberg limit could persist for large $N$ in some special measurement scenarios \cite{chen2018achieving,chen2018heisenberg}, a simple way to scale up quantum metrology advantages is currently missing.  

Recently, research in the foundations of quantum mechanics identified a new quantum resource, namely, the ability to perform operations in an indefinite causal order (ICO)  \cite{hardy2007towards,chiribella2009beyond,oreshkov2012quantum,chiribella2013quantum}.     A particular form of ICO, known as the quantum SWITCH \cite{chiribella2009beyond,chiribella2013quantum}, provides coherent control over the order of operations and has been found to have applications in various areas of quantum information\cite{chiribella2012perfect,procopio2015experimental,colnaghi2012quantum,araujo2014computational,renner2022computational,ebler2018enhanced,goswami2020increasing,chiribella2021quantum,guo2020experimental,felce2020quantum,zhu2023charging}, including quantum metrology~\cite{zhao2020quantum,chapeau2021noisy,an2024noisy,yin2023experimental}. In particular, the quantum SWITCH was shown to offer a major scaling advantage in the measurement of a geometric phase associated with two groups of $N$ phase-space displacements \cite{zhao2020quantum,yin2023experimental}. 

Although ICO offers intriguing advantages in quantum metrology,  these advantages have been so far limited to rather specific problems. Notably, no application has been found for practically relevant problems such as phase estimation or, more generally, the estimation of a single parameter associated with quantum dynamics.  Exploiting  ICO in the single-parameter scenario appears far from straightforward, especially in the ideal case of unitary dynamics with time-dependent Hamiltonians, because the unitary evolution commutes at all times, and therefore the order in which different snapshots of dynamical evolution are executed does not affect the final state of the probe.   A possible way around this is to combine the probed dynamics with some known operations and to combine these operations in an ICO.  While these additional operations can always be regarded as part of the setup, the use of ICO between known operations and unknown dynamics can offer practical benefits, such as enabling non-classical precision scalings with easy-to-prepare probe states.

In this work, we demonstrate a practical application of ICO in the estimation of an unknown rotation angle, imprinted on a vortex beam subject to operations in a coherently controlled order. The eigenvalue spectrum of the vortex beam can be expanded according to the topological charge of the photon. Recent advances have shown values of the orbital angular momentum (OAM) up to $10010\hbar$ \cite{fickler2016quantum}, which is naturally appropriate to measure rotation angle \cite{wang2021application,xia2024nanoradian,d2013photonic,courtial1998measurement} and rotation angular velocity \cite{barreiro2006spectroscopic,zhang2018free}. A hybrid setup inspired by the quantum SWITCH is built to probe a rotation by an angle $2m\theta$, by combining it with operations that increase or decrease the OAM by $l\hbar$, which are used in a coherently controlled order as leverage to enhance the estimation of the unknown angle $\theta$. By using these leverage operations in an ICO,  we achieve a precision enhancement by a factor of $4ml$,  which we experimentally demonstrate up to the value 4096 by setting $l=128$ and $m=8$.  By taking noise into account, the overall enhancement can maintain a high factor of 2317, and the ultimate precision in our experiment is $0.0105^{\prime \prime}$ when using $7.16\times10^7$ photons, corresponding to a normalized precision of $4.3\times10^{-4}$rad for one photon.

\section{Results}

\begin{figure}
    	\centering
	\includegraphics[width=0.8\linewidth]{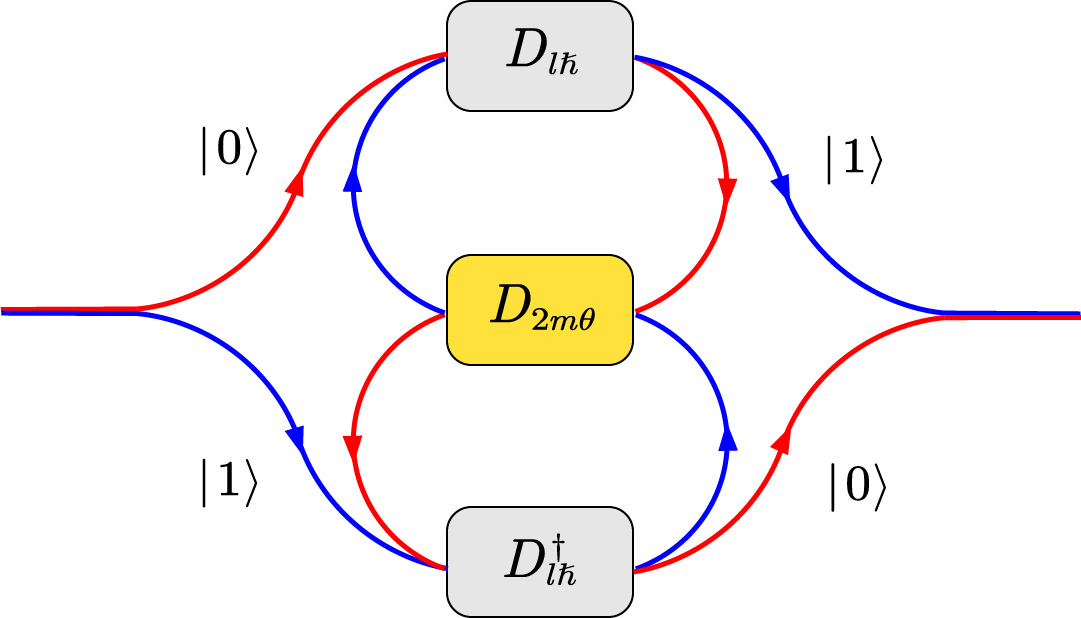}
	\caption{\textbf{Quantum SWITCH setup for the measurement of rotation angle. }   A rotation of the orbital angular momentum, denoted by $D_{2m\theta}$, is performed between an operation $D_{l\hbar}$ that increases the orbital angular momentum and its inversion $D^{\dagger}_{l\hbar}$, with the order of these two operations coherently controlled by the state of a qubit.   
 }
 \label{fig1}
\end{figure}

\begin{figure*}
	\centering
	\includegraphics[width=0.95\linewidth]{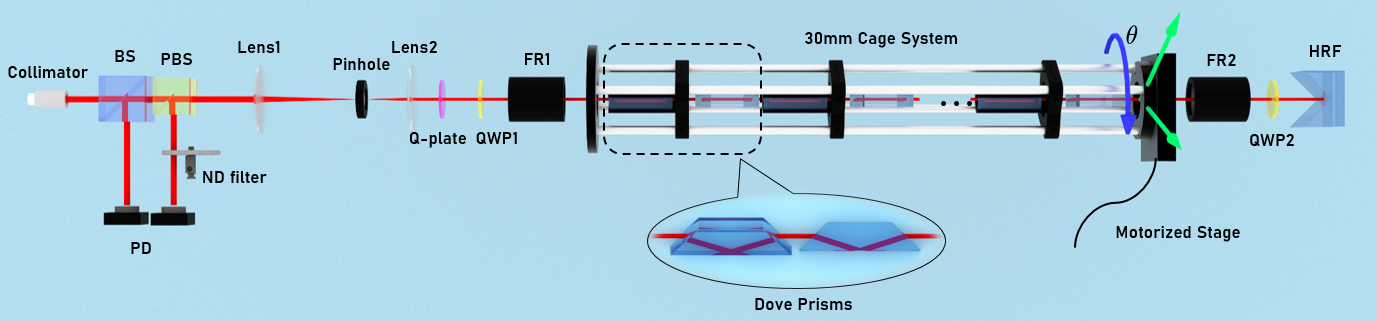}
	\caption{\textbf{Experimental setup of the hybrid quantum SWITCH for the measurement of a rotation angle.}  Our setup uses the photon polarization to control the order of two opposite shifts of the orbital angular momentum. These shifts are performed before and after an unknown transversal rotation, reproducing the same overall dynamics arising from the quantum SWITCH of discrete and continuous variables.    Firstly, half of the photons from a DFB laser transmit the beam splitter (BS) and become horizontally polarized after passing the polarization beam splitter (PBS), which can be written into the superposition of left- and right-handed circular polarization as $\frac{1}{\sqrt{2}}(\lvert R\rangle+\lvert L\rangle)$ and serves as the control qubit. A Q-plate couples the spin-orbital momentum to create a joint quantum state as $\frac{1}{\sqrt{2}}( \lvert R\rangle \lvert +l \rangle + \lvert L\rangle \lvert -l\rangle)$ and thus introduces $2l\hbar$ relative OAM displacement between the two orders.
  After that, a $0^{\circ}$ quarter wave plate (QWP) and a $\pi/4$ Faraday rotator (FR), namely, the QWP1-FR1 suite transforms circular polarization to linear polarization ($\lvert R \rangle \rightarrow \lvert V\rangle$ and $\lvert L \rangle \rightarrow \lvert H\rangle$), and then $m/2$ pairs of Dove prisms rotate the transversal mode for both orders to encode $\theta$ to the joint state $\frac{1}{\sqrt{2}}( e^{-iml\theta}\lvert V\rangle \lvert +l\rangle + e^{iml\theta}\lvert H\rangle \lvert -l\rangle)$. Each pair of Dove prisms includes a stationary prism and the other rotatable prism as shown in the inset. A $\pi/4$ FR and a following $0^{\circ}$ QWP (FR2-QWP2) conduct the linear-to-circular transformation of the photon polarization, and the ending hollow roof prism  (HRP) flips the circular polarization without inversion of OAM. The reflected photons exchange the linear polarization ($\lvert V \rangle \rightarrow -\lvert H\rangle$ and $\lvert H \rangle \rightarrow \lvert V\rangle$) after repassing the FR2-QWP2 suite; and thus, the polarization deflection due to the internal reflection on the prisms is eliminated and the control qubit is protected to proceed the quantum SWITCH.  After reversely passing the Dove prisms, the photons undergo twifold rotating operations and joint state changes to $\frac{1}{\sqrt{2}}( -e^{-2iml\theta}\lvert H\rangle \lvert +l\rangle + e^{2iml\theta}\lvert V\rangle \lvert -l\rangle)$. Then the FR1-QWP1 suite retrieves the linear polarization to circular polarization, and the joint state evolves to  $\frac{1}{\sqrt{2}}( -e^{-2iml\theta}\lvert R\rangle \lvert +l\rangle + e^{2iml\theta}\lvert L\rangle \lvert -l\rangle)$. Once again the Q-plate introduces the spin-orbital coupling but oppositely for the two orders, and the outcome state from the quantum SWITCH is $\frac{1}{\sqrt{2}}(e^{-2iml \theta}\lvert L\rangle+e^{2iml \theta} \lvert R\rangle )$, which acquires a geometric phase of $4ml\theta$ but no topological charge of OAM. The transverse wavefunction $\lvert\Phi \rangle$ undergoes $2\theta$ rotation and is not measured.  After being spatially filtered by the pinhole placed on the common focal plane of Lens1 and 2, photons exiting from the reflection ports of the PBS and BS are recorded by two photodiodes (PDs) to calculate the projecting probability $P(\theta)$, and the ND filter is used to balance the collection efficiency between the two ports. 
 }
  \label{fig2}
\end{figure*}

Here we provide a setup for the estimation of an unknown rotation angle $
\theta$ imprinted on the OAM degrees of freedom. These degrees of freedom are characterized by two quantum-mechanical observables, namely the azimuth angle $\hat{\theta}$ and the  $z$-component of the orbital angular momentum $\hat{L_z}$. The commutation between $\hat{\theta}$ and $\hat{L_z}$ manifests a complex expression  due to the imposition of periodic boundary conditions \cite{judge1964uncertainty,pegg2005minimum}. For simplicity and consistency with our experimental setting, we constraint the rotation angle as $\theta\in(-\pi, \pi)$. Consequently, the two observables adhere to the canonical commutation relation $[\hat{\theta},\hat{L}_z] =i\hbar$, which parallels the the commutation relation between position and momentum. An important difference here is that the spectrum of the orbital angular momentum is discrete. Physically, the OAM is not a continuous degree of freedom because only the Laguerre-Gaussian beam with an integer topological charge is the eigenmode of propagation. Thanks to the discreteness of the angular momentum,   displacements in this degree of freedom can be precisely implemented,  thus providing a natural candidate for the leverage operation in our ICO setup.  This situation is different from the case of continuous observables, like position and momentum,  for which precise control of displacements is more challenging.   

Our setup probes an unknown rotation $D_{2m\theta}=e^{-\frac{2i}{\hbar}m\hat{L_z}\theta}$ by sandwiching it between the discrete operations $D_{\pm l\hbar}=e^{\pm i l \hat{\theta}}$ and $D_{\pm l\hbar}^{\dagger}$, which shifts the $z$-component angular momentum upwards and downwards, respectively, by $l\hbar$. The order of the two shifts is coherently controlled by a qubit, as illustrated in Figure.~\ref{fig1}, giving rise to the overall evolution  $W:=D_{l\hbar}^{\dagger}D_{2m\theta}D_{l\hbar}\otimes\lvert 0\rangle \langle 0\rvert+D_{l\hbar}D_{2m\theta}D_{l\hbar}^{\dagger}\otimes \lvert 1\rangle \langle 1\rvert$.  The operator $W$ is unitarily equivalent to the operator 
\begin{equation}
    W_{\rm QS}=D_{2m\theta} D_{2l\hbar} \otimes \lvert 0\rangle \langle 0\rvert +D_{2l\hbar} D_{2m\theta} \otimes \lvert 1\rangle \langle 1\rvert, 
\end{equation}
which reproduces the same evolution that would arise if the gates $D_{2m\theta}$ and $D_{2l\hbar}$ were combined in the quantum SWITCH.  With a little abuse of terminology, in the following we will refer to our setup as a hybrid quantum SWITCH setup, the term ``hybrid'' referring to the presence of both discrete and continuous variables.    

Since the observables $\hat{\theta}$ and   $\hat{L_z}$ are canonically conjugate, the corresponding displacement satisfies the relation  $D_{2l\hbar} D_{2m\theta}   =  e^{i 4ml \theta}D_{2m\theta} D_{2l\hbar}$, which implies $W_{\rm QS}  =D_{2m\theta} D_{2l\hbar} \otimes U_{4ml \theta}$, with $U_{4ml \theta}  : =  |0\rangle\langle 0|  + e^{i 4ml \theta}  \, |1\rangle\langle 1|$.  In other words, the phase $\Psi  =  4ml\theta$ is transferred from the OAM to the control qubit, in the same way as in the geometric phase setup in Refs.  \cite{zhao2020quantum,yin2023experimental}.  
  By measuring the control qubit, it is then possible to estimate the phase  $\Psi$ and to compute $\theta$ as $\theta=\Psi/4ml$. Crucially, the factor $4ml$ offers a nonlinear precision enhancement while merely consuming linearly increasing resources, i.e., totally, $N_{g}=2(m+l)$ gates are called inside the quantum SWITCH. This nonlinear scaling dramatically surpasses the linear Heisenberg limit as $1/N_{g}$, which has been traditionally regarded as the ultimate limit imposed by the Heisenberg uncertainty principle. However, when encoding a parameter $\theta$ via a rotating gate $U_{\theta}$, the HUP does not pose an absolute constraint on the vanishing speed of $\delta \theta$ with increasing probe particles or evolution gates, nevertheless the parameter-based HUP expressed as $\delta \theta \cdot \Delta h \geq \frac{1}{2}$ has to be satisfied \cite{hou2021super}. In this trade-off relation $\Delta h$  represents the standard deviation (SD) of infinitesimal translation generator $\hat{h}$ defined as $\hat{h}=i[\partial_{\theta}U_{\theta}] U_{\theta}^{\dagger}$. In this sense, increasing $\Delta h$ could minimize the uncertainty in estimating $\theta$.

For a conventional scheme of multi-pass encoding, e.g., applying $m$ rotating gates to a  $l$-level OAM beam results in a generator $h_{MP}=i[\partial_{\theta}(U_{\theta}^{(m)})](U_{\theta}^{(m)})^{\dagger}=\frac{2m\hat{L}_z}{\hbar}$, of which the SD is independent of $l$ and calculated as $\Delta h_{MP}=2m\Delta L_z$, with $\Delta L_z$ representing the average uncertainty of the initial OAM state ((see Supplementary Material Sec. I for detailed calculation)). 

In our quantum SWITCH scheme, the SD of the generator $h_{QS}=i[\partial_{\theta}W_{QS}]W_{QS}^{\dagger}$ is calculated as $\Delta h_{QS}=2m\frac{\Delta L_z}{\hbar}+2ml$, in which the nonlinear item $2ml$ arises with the factors $m$ and $l$ both representing the gate query times in encoding (see Supplementary Material Sec. I for detailed calculation). Accordingly, $\delta\theta$ can vanish with a nonlinear scaling as $\frac{1}{ml}$ while keeping the parameter-based HUP hold. This result is consistent with our previous work in which the gate query times are both $N$ and hence the precision attains the scaling of $\frac{1}{N^2}$ \cite{yin2023experimental}.

In our experiment, the hybrid quantum SWITCH is constructed by using the photon polarization as the control qubit to realize the superposition of two alternative orders. The coherent control arises from the spin-orbital coupling \cite{marrucci2006optical} introduced by a Q-plate, which couples the OAM and polarization degrees of freedom as
\begin{equation}
\begin{aligned}
Q_{l}&=D_{l\hbar}\otimes \lvert R\rangle \langle L \rvert+ D_{l\hbar}^{\dagger}\otimes \lvert L\rangle \langle R \rvert,
\end{aligned}
\end{equation}
where $\lvert L\rangle$ and $\lvert R\rangle$ refer to the left- and right-handed circular polarized states. After passing the Q-plate, the $\lvert L\rangle$ and $\lvert R\rangle$ components of a linear polarized photon experience opposite OAM operations as $\pm l\hbar$, with the relative OAM displacement identified to be $2l\hbar$.

The unknown rotation operations are implemented through $m/2$ (m is an even number) pairs of Dove prisms, and each pair consists of one rotatable prism aligned with a relative angle  $\theta$ to the other stationary prism (see Methods for details). The transverse mode of photons rotates $2\theta$ after passing each pair of prisms. 

The setup is shown in Fig. 2, the quantum SWITCH in our experiment is realized through a co-linear and round-trip interferometer, which is distinct from many previous apparatuses, of which the two orders are subject to two spatially separated arms \cite{guo2020experimental,procopio2015experimental,yin2023experimental}. The co-linear feature avoids phase fluctuation due to the different circumstances surrounding two separated arms, and the round-trip configuration doubles the number of rotation operations to acquire an additional two-fold enhancement. One obstacle to adopting the round-trip configuration is that a reflecting mirror flips the parity of transverse mode, which is then rotated back to the initial orientation after double passing the Dove prisms, and no geometric phase can be accumulated.  To tackle this issue, an HRP is used instead of a mirror as the ending reflection element. After being twice reflected on the HRP, the OAM mode remains unchanged, and the rotation operation reserves, which assure the generation of the geometric phase in the quantum SWITCH (see Supplementary Material Sec. II for details). To implement the quantum SWITCH, the control qubit has to be well maintained. However, the internal total reflection of the Dove prisms incurs a minimal polarization deflection hinging on the unknown parameter $\theta$, which undermines the interference visibility of the two orders. In our scheme, this deflection effect is eliminated by flipping the photon polarization after double passing the FR2-QWP2 suite before the HRP (see Methods for details).

\begin{figure*}[t]
	\centering
	\subfigure[m=2, l=1]{\label{fig4a}
    \includegraphics[width=0.3\linewidth]{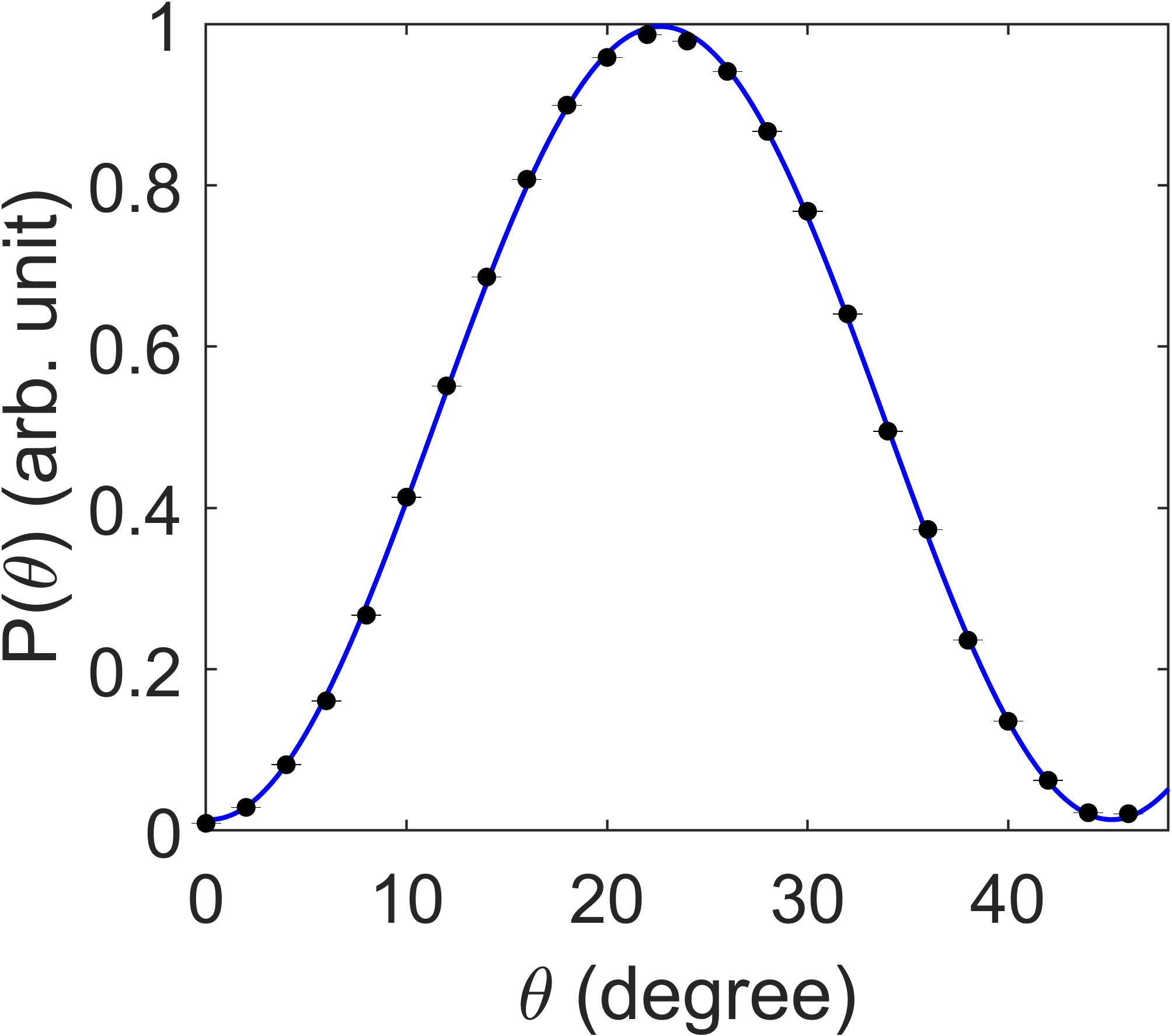}}
    \subfigure[m=4, l=2]{\label{fig4b}
    \includegraphics[width=0.3\linewidth]{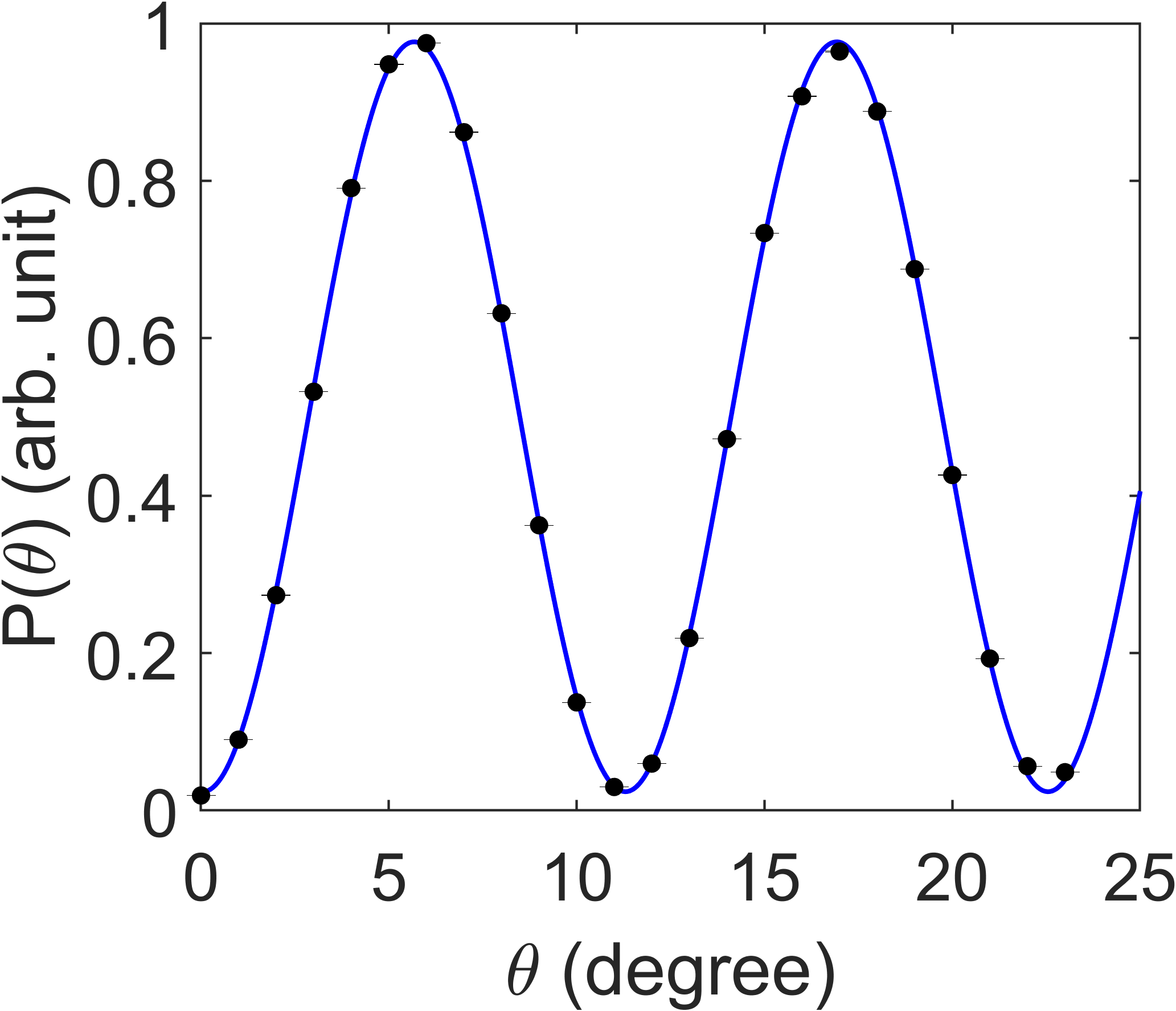}}
    \subfigure[m=6, l=3]{\label{fig4c}
    \includegraphics[width=0.3\linewidth]{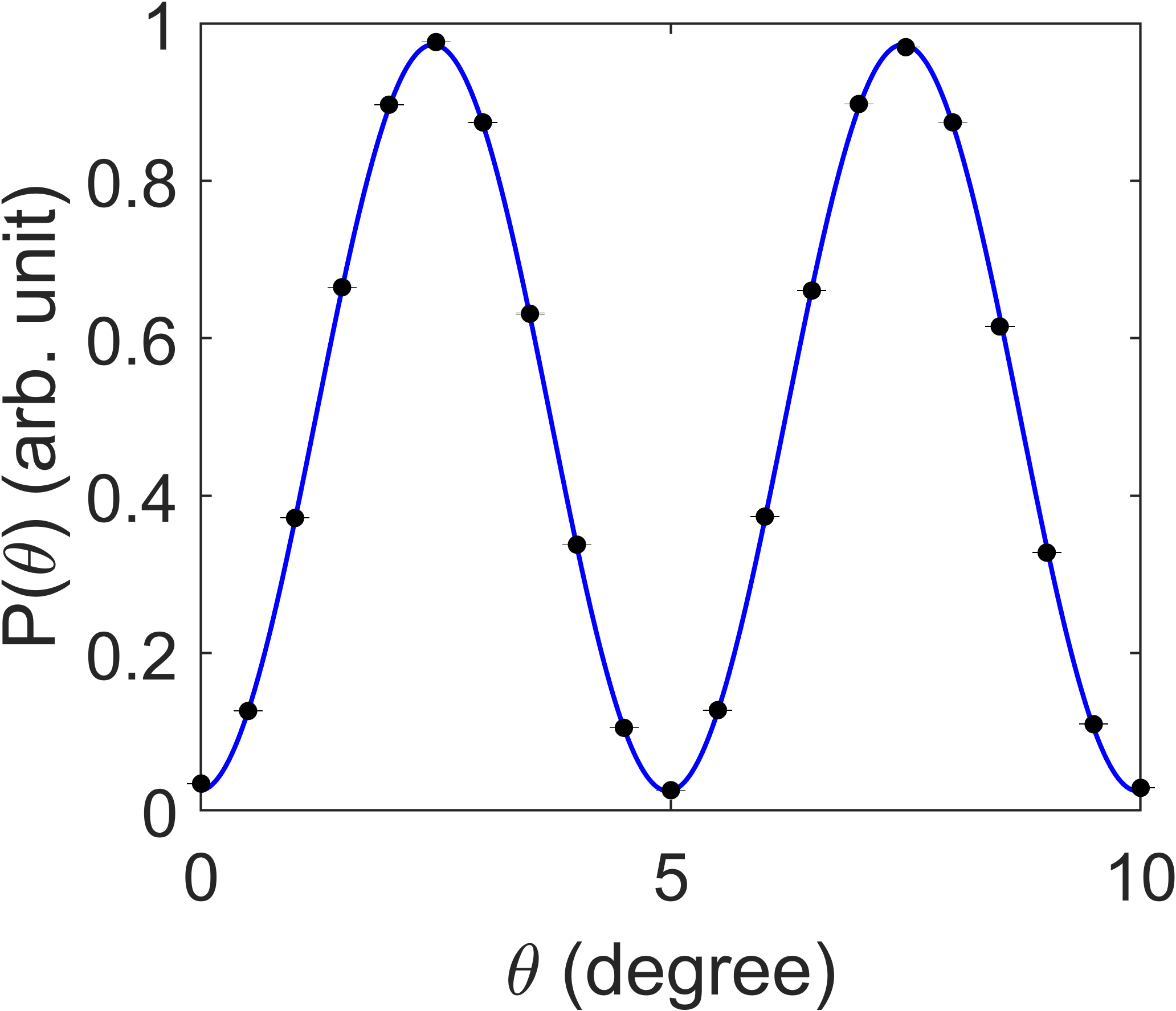}}
    \subfigure[m=8, l=4]{\label{fig4d}
    \includegraphics[width=0.3\linewidth]{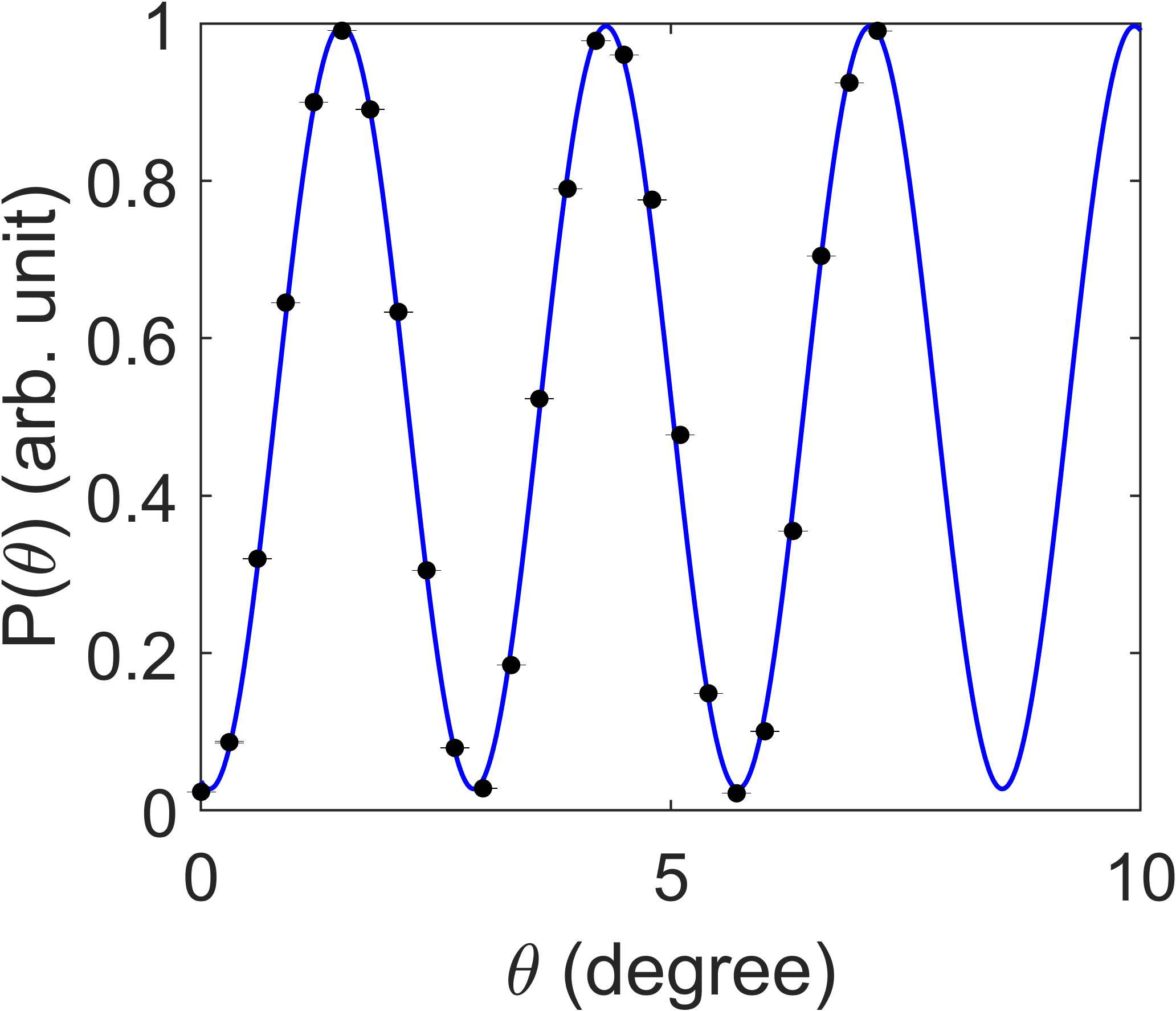}}
    \subfigure[m=12, l=6]{\label{fig4e}
    \includegraphics[width=0.3\linewidth]{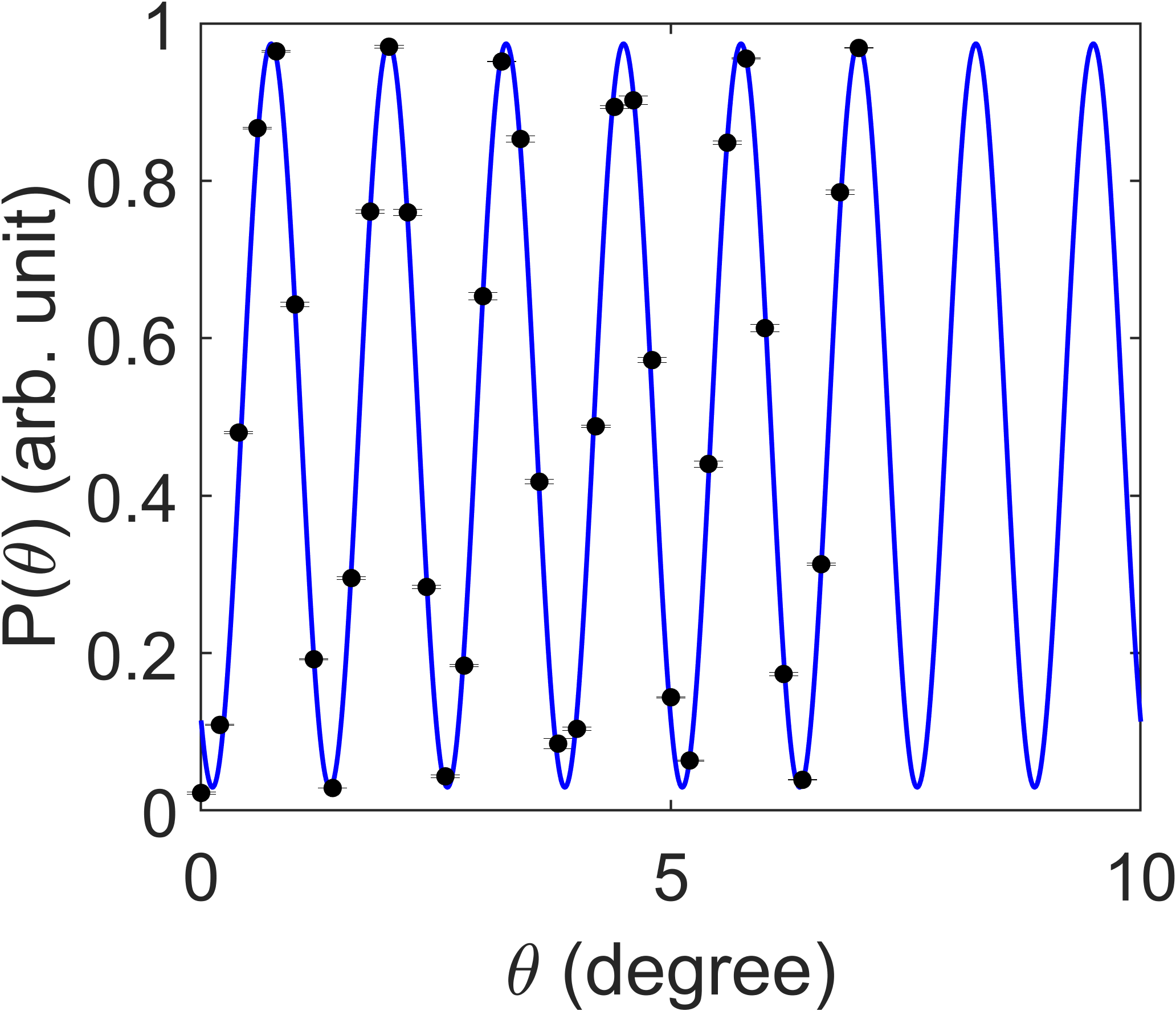}}
    \subfigure[m=8, l=128 at a range of 1 degree]{\label{fig4f}
    \includegraphics[width=0.3\linewidth]{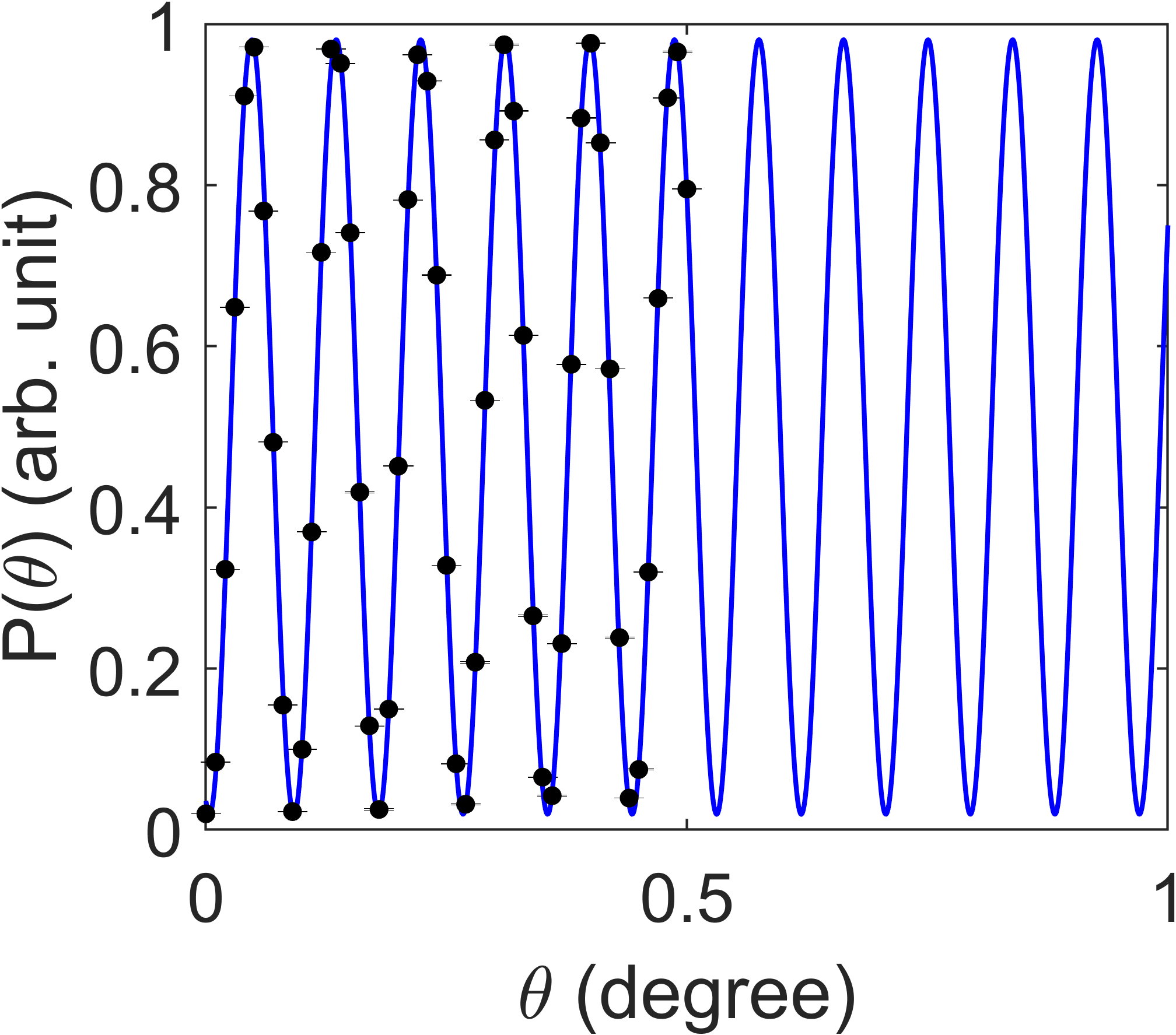}}
	\caption{\textbf{The interference fringe derived from the geometric phase on the control qubit.} The projection probability $P(\theta)$ is measured by changing $\theta$, for which the data points are fitted with the function given by Eq.~\eqref{eq4}, and the fitting curves are shown in blue. A total of six combinations of $m,l$ are investigated, and the period reducing with the scaling of $1/ml$ indicates the nonlinearly enhanced precision to measure $\theta$.
 In each trial of the experiment, $P(\theta)$ is calculated from the results of projective measurement on $\nu=7\times10^7$ photons, and the data points denote the mean values ($\pm$ RMSE) of 60 trials of the experiment.
As we can see, all the data points fall at the fitting curve and the error-bar is not visible since the error is smaller than the dot size.}
 \label{fig3}
 \end{figure*}

After passing the round-trip including a $l$-order Q-plate and $m$ Dove prisms, a horizontally polarized photon experiences $2m\theta$ angular displacement and $2l\hbar$ relative OAM displacement, in the superposition of two alternative orders. The control-system joint state evolves as
\begin{equation}
\begin{aligned}
	 &Q_{l}  D_{2m\theta} Q_{l}\lvert H\rangle \lvert\Phi\rangle
	 \\&=\frac{1}{\sqrt{2}} (D_{l\hbar}^{\dagger} D_{2m\theta} D_{l\hbar} \lvert\Phi\rangle \otimes \lvert L\rangle+ D_{l\hbar}D_{2m\theta} D_{l\hbar}^{\dagger} \lvert\Phi\rangle \otimes \lvert R\rangle)
  \\& = \frac{1}{\sqrt{2}}(e^{-2iml \theta}\lvert L\rangle+e^{2iml \theta} \lvert R\rangle )\otimes  D_{2m\theta} \lvert \Phi \rangle,
\end{aligned}
\end{equation}
where $\lvert H\rangle$ denotes the horizontally polarized state, and $\lvert \Phi\rangle$ denotes the spatial wave function of photons. In this process, $D_{l\hbar}$ and $D_{l\hbar}^{\dagger}$ are exchanged in different causal orders.

The final estimation of $\theta$ is achieved by projecting the control qubit to    $\lvert V\rangle \langle V\rvert$, of which the probability should be calculated as
\begin{equation}
    P(\theta)=\frac{1}{2}\left[ 1-\cos(4ml \theta+\phi_0)\right] ,
    \label{eq4}
\end{equation} 
where $\phi_0$ represents an initial phase.
The value of $\theta$ can be directly extrapolated from the projective measurement on the control qubit, while the spatial mode is completely discarded to avoid the complicated measurement of the spatial distribution which may introduce more technical noise. Most of all, the factor $4m l\theta$ indicates a non-linear quantum enhancement of the measurement precision of $\theta$ while no classical non-linearity is involved.

To prove the quantum nonlinear enhancement, various values of $m$ and $l$ are tested in our experiment and Fig. 3 shows the interference fringe predicted by Eq.~\eqref{eq4}. It can be seen that for each pairing of $(m,l)$, the measured projecting probabilities drop on the theoretical curve.

A theoretical analysis shows that the Fisher information of our scheme is calculated as $16\nu m^2 l^2$ (see Supplementary Material Sec. III for details), and the theoretical precision to estimate $\theta$ can then be obtained from the Cram\'er-Rao bound and written as
\begin{equation}
    \delta \theta \geq \frac{1}{4\sqrt{\nu} ml},
    \label{eq5}
\end{equation}
where $\nu$ is the photon number used in each trial of measurement.

\begin{figure}
	\centering
    \includegraphics[width=0.8\linewidth]{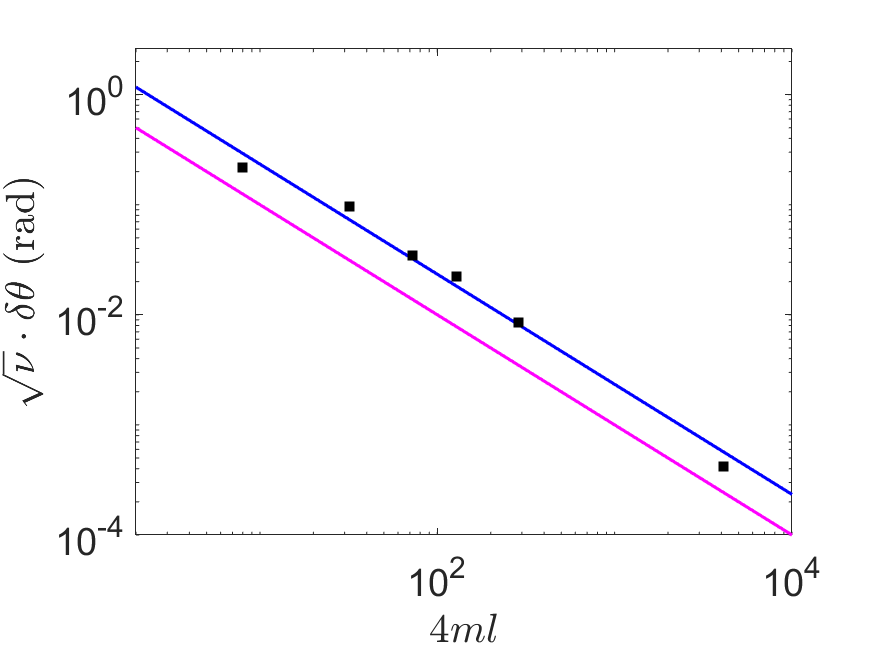}
	\caption{\textbf{Nonlinear scaling of the measurement precision normalized for one photon}.  The black data points are the measured RMSE $\delta \theta$ normalized by the square root of photon number $ \nu$, which represents the measurement precision achieved with one photon for varying $4ml$. The blue fitting line is close to the theoretical Cram\'er-Rao bound  $\delta \theta \sqrt{\nu}  = \frac{1}{4ml}$ (red line), with a gap factor of 2.4.  }
\label{fig4}
 \end{figure}

For each pairing of ($m, l$), $P(\theta)$ is measured by using approximately $7\times10^7$ photons to get one estimation of $\theta$ from Eq.~\eqref{eq4}. The true value and uncertainty of $\theta$ are identified as the mean and root mean square error (RMSE) of 60 trials of estimation. 
The RMSE with various $4ml$ is given in Fig.~\ref{fig4}, which approaches the Cram\'er-Rao bound and vanishes with the speed $1/4ml$ as predicted by Eq.~\eqref{eq5}. The experimental fitting curve is only worse than the Cram\'er-Rao bound by a factor approximately of 2.5, which is mainly due to the noise when mechanical rolling Dove prisms. The ultimate precision is obtained by using a 128-order Q-plate and 4 pairs of Dove prisms, which jointly promise an enhanced factor as high as 4096 and practically achieve 2317 in the experiment. When measuring a rotating angle of 0.025 degrees with $7.16\times 10^7$ photons, the absolute precision is up to $0.0105^{\prime \prime}$ corresponding to a normalized precision of $4.3\times 10^{-4}$ rad for one photon. Both the enhancement factor and the normalized precision achieved in this study significantly surpass those obtained in the indefinite evolution scheme that employs a reversing operation on OAM beams \cite{xia2024nanoradian}, as well as those in the classical scheme named photonic gear \cite{d2013photonic}, which aims to measure the relative rotation between two reference frames responsible for state preparation and measurement, respectively. By contrast, our proposed scheme addresses more general scenarios where a rotating object is completely isolated from the entire apparatus.

\section{Discussion}
Classical nonlinear interactions among $N$ particles are known to lead to a nonlinear scaling of the measurement uncertainty as $\Delta \propto 1/N^{\frac{3}{2}}$ \cite{boixo2007generalized,roy2008exponentially,choi2008bose,napolitano2011interaction}.    However, such enhancements typically do not apply to the basic scenario of $N$ independent and identical processes,  and have been subject to dispute  \cite{zwierz2012ultimate,berry2012optimal,hall2012does}; especially, nonlinear interactions have been found to be unable to break the classical limit in terms of relevant resources such as the energy of the probes {\cite{hall2012does, berry2012optimal} and universal resource count \cite{zwierz2010general,zwierz2012ultimate}. 
 In stark contrast, our scheme only harnesses non-interacting photons and independent processes, and the nonlinear enhancement definitely arises from the quantum superposition between two alternative orders.  This approach opens up a route to create quantum nonlinear enhancement by incorporating ICO evolution.

Our scheme is fundamentally different from conventional quantum metrology schemes, in which one has to prepare the probe into a known state, then couple it with the system to draw the target parameter into the joint quantum state, and finally apply a certain measurement to the probe. In contrast, our scheme requires no specially tailored input state; even a scrambled speckle can result in the same results. Instead, our scheme exploits the ability to coherently control leverage operations performed before and after the target parameter is encoded on the system.  Physically, the difference is clear from the fact that the order of the Q-plates, and not the specific OAM state, is important in our setup.   The final measurement we apply is a projective measurement on the control qubit, and the distribution in two conjugate observables $\hat{\theta}$ and $\hat{L}_z$ are discarded.

Increasing $\Delta L_z$ could yield a large SD of the generator exampled by $\Delta h_{MP}$, and then boost the quantum Fisher information. However, the optimal measurement is formidably hard since it requires projective measurement to a superposition state of two OAM eigenmodes without photon loss \cite{bolduc2013exact,bent2015experimental}(see Supplementary material Sec. III for detailed FI analysis).  In contrast, ICO encodes the parameter to the relative phase of the control qubit, and the optimal projective measurement can be simply executed with a polarizer. Applying multi-pass rotating gates in a Mach-Zehnder interferometer (MZI)-based scheme \cite{courtial1998measurement} renders QFI only one quarter one-sixteenth of that in our scheme, furthermore, this scheme fails to provide a practically competitive level of precision as it is afflicted with the inherent rotational symmetry in the input beam distribution and the path deviation of the MZI. (see Supplementary material Sec. III for detailed analysis)

From a technical point of view, our work provides a method for experimentally constructing a photonic quantum SWITCH not involving spatially separated arms. The control operation relies on the polarization of photons so that the two inverse orders can be coherently combined in one co-linear path and the phase offset can be eliminated by a QWP.  Moreover, the polarization deflection can also be overcome by a round-trip configuration. 

Considering the quantum resources are defined as the number of used gates in the evolution, the power of our scheme can be characterized as the utilization efficiency of these resources; concretely, it is quantified as the normalized precision achieved by one photon, which attains quantum nonlinear scaling of $1/4ml$ and achieves $4.3\times10^{-4}$rad in the rotation angle measurement. The overall RMSE reduces with the increasing photon number $\nu$, while merely showing a classical scaling of $1/\sqrt{\nu}$. 

Our scheme, experimentally demonstrated for rotations on an OAM system, can in principle be applied to the problem of phase estimation, or the estimation of other shift parameters associated with a generator with a discrete spectrum. Overall, the framework and experimental techniques developed here open up new opportunities to advance the frontier of precision measurements in quantum technologies.

\bibliographystyle{naturemag.bst}
\expandafter\ifx\csname url\endcsname\relax
  \def\url#1{\texttt{#1}}\fi
\expandafter\ifx\csname urlprefix\endcsname\relax\def\urlprefix{URL }\fi
\bibliography{ref.bib}

\section{Methods}

\subsection{Details of the rotating system}
  To transmit OAM beam below 128 order, Dove prisms with $5mm\times 5mm$  clear aperture are used.  All the rotatable Dove prisms are aligned with an established line by a $30mm$ cage system, which can be rotated through a motorized rotation stage.  The initial angle of Dove prisms does not influence the result since it can always be made zero through QWP1. The visibility of interference between two circular polarized states is $\sim 0.96$ after being improved by a spatial filter system consisting of two lenses and a pinhole. The focus lengths of Lens1 and Lens2 are respectively $f=10cm$ lens and $f=5cm$, which are used to focus and collimate the beam. The pinhole is placed in the common focus of the lenses to block the stray light due to the unflatness of the reflecting surface. For a 128-order OAM beam suffering a large radius and divergence angle, we exchange the position of Lens1 and Lens2 to expand the incident beam and change the pinhole size to $200\mu m$. The Dove prism size is changed to $10mm \times 10mm$ to accommodate the large profile of OAM beam.

\subsection{Protecting the control qubit against polarization deflection on prisms}

Dove prism introduces an extra phase and amplitude difference between the s-wave and the p-wave when the total internal reflection occurs on the bottom surface. This polarization-dependent transmission necessarily changes the polarization state of photons which serves as the control qubit of the quantum SWITCH, and diminishes the interference visibility between the two orders. 
This polarization deflection cannot be pre- or post-compensated since the Jones matrix of Dove prism $J(\alpha)$ also hinges on the unknown $\theta$. We develop a method to eliminate this deflection by exploiting the relation that $J(\alpha)J(\alpha+\pi/2)$ is proportional to a unit matrix $I$  (see Supplementary Material Sec. IV for details), where $\alpha$ is the angle between the axis of Dove prism and polarization of photons. This relation suggests that any depolarization can be eliminated after double-passing the prism with orthogonal polarization. 

The polarization flipping suite can be realized through a $0^{\circ}$ aligned QWP and a following FR rotating the photon polarization by 45 $^{\circ}$. Double-pass of one such suite before the ending HRP flips the photon polarization; and thus the Jones matrix of $m$ pairs of Dove prism can be written as $J_{m}(\theta)  \dots J_2(\theta) J_1(\theta)$ and $J_{1}(\theta+\pi/2) J_{2}(\theta+\pi/2) \dots J_{m}(\theta+\pi/2)$ for the forth and back trip respectively, which jointly result in an identity.

\section{Data Availability}
The datasets supporting the findings of this work are available from the corresponding authors upon reasonable request.

\section{Acknowledgement}

We thank Professor Zhibo Hou for the helpful discussion on theory. This work was supported by the Innovation Program for Quantum Science and Technology (Nos. 2021ZD0301200, 2021ZD0301400, 2021ZD0302000), National Natural Science Foundation of China (Grant Nos. 12122410, 12350006), Anhui Initiative in Quantum Information Technologies (AHY060300), the National Natural Science Foundation of China via Excellent Young Scientists Fund (Hong Kong and Macau) Project 12322516, the Hong Kong Research Grant Council (RGC) through the Early Career Scheme (ECS) grant 27310822 and the General Research Fund (GRF) grant 17303923, the Fundamental Research Funds for the Central Universities (Grant No. YD2030002026), by the Hong Kong Research Grant Council through grants T45-406/23-R and R7035-21F, by the  Ministry of Science and Technology of China through grant 2023ZD0300600, and by the John Templeton Foundation through grant 62312, The Quantum Information Structure of Spacetime (qiss.fr).
The opinions expressed in this publication are those of the authors and do not necessarily reflect the views of the John Templeton Foundation. Research at the Perimeter Institute is supported by the Government of Canada through the Department of Innovation, Science and Economic Development Canada and by the Province of Ontario through the Ministry of Research, Innovation and Science.

\section{Competing interests}
The authors declare no competing interests.

\end{document}